\documentclass[12pt,preprint]{aastex}

\newcommand{\Vcom}{V_0}
\newcommand{\Vcirc}{V_{\rm circ}}
\newcommand{\Vend}{V_{\rm end}}
\newcommand{\Vpred}{V_{\rm pred}}
\newcommand{\Rend}{R_{\rm end}}
\newcommand{\Rexp}{R_{\rm exp}}
\newcommand{\kms}{\,{\rm km}\,{\rm s}^{-1}}
\newcommand{\mpc}{\,{\rm Mpc}}
\newcommand{\hubunits}{\,{\rm km}\,{\rm s}^{-1}\,{\rm Mpc}^{-1}}
\newcommand{\Mr}{M_r}
\newcommand{\halpha}{H$\alpha$}
\newcommand{\Vmax}{V_{\rm max}}

\slugcomment{Submitted to ApJL}

\shorttitle{Rotation Velocities of Faint Galaxies}
\shortauthors{Pizagno et al.}

\begin{document}

\title{Rotation Velocities of Two Low Luminosity Field Galaxies}

\author{James Pizagno\altaffilmark{1}, Michael R. Blanton\altaffilmark{2}, 
David H. Weinberg\altaffilmark{1}, Neta A. Bahcall\altaffilmark{3}, Jon 
Brinkmann\altaffilmark{4} }

\altaffiltext{1}{Department of Astronomy, Ohio State University, Columbus, OH 
43210}
\altaffiltext{2}{Department of Physics, New York University, New York, NY 
10003}
\altaffiltext{3}{Princeton University Observatory, Princeton, NJ 08544}
\altaffiltext{4}{Apache Point Observatory, 2001 Apache Point Road, P.O. Box 59, Sunspot, NM 88349-0059}

\begin{abstract}
We present \halpha\ rotation curves of two low luminosity field galaxies with
$r$-band absolute magnitudes $\Mr=-13.9^{+0.8}_{-0.5}$ and 
$\Mr=-14.7^{+0.3}_{-0.2}$ (for $h\equiv H_0/100\hubunits =0.7$; the large 
error bars reflect distance uncertainties). Most previously studied galaxies
in this luminosity range are members of groups defined by brighter galaxies, 
but these two systems, selected from Blanton et al.'s (2004) sample of low
luminosity galaxies in the Sloan Digital Sky Survey (SDSS), appear to have no
bright companions. The measured rotation speeds at the outer extent of the
\halpha\ rotation curves, $34.8\pm 3.8\kms$ and $30.9\pm 7.2\kms$, are larger
than the values of $16.0^{+6.1}_{-5.4}\kms$ and $20.9^{+6.2}_{-5.2}\kms$
predicted by extrapolating the inverse Tully-Fisher relation of luminous 
SDSS galaxies to these faint luminosities. However, a previous HI measurement
of the first galaxy shows that it has a gas mass similar to its stellar mass,
and the total baryonic mass is consistent with that predicted by McGaugh 
et al.'s (2000) ``baryonic Tully-Fisher relation.'' We find $r$-band dynamical 
mass-to-light ratios within the radii of the last \halpha\ data points (about
1.8 disk scale lengths in each case) of $12.6^{+4.7}_{-4.5}\, M_\odot/L_\odot$
and $4.8_{-2.1}^{+2.5}\, M_\odot/L_\odot$, much higher than the values 
$\sim 1\,M_\odot/L_\odot$ expected for the stellar populations. The dynamical
properties of these galaxies, including the rotation speeds and evidence
for high gas fractions and dark matter domination within the luminous extent
of the galaxy, are consistent with those of previously studied faint galaxies
in nearby groups. Further studies of the SDSS sample will allow characterization
of low luminosity galaxies over the full range of environments in which they
reside.
\end{abstract}


\section{Introduction}
The Sloan Digital Sky Survey \citep{york00} galaxy redshift 
survey has an unprecedented combination of large area, 
depth, and photometric quality, thanks to the combination 
of a large format camera \citep{gunn98}, high throughput 
multi-object spectrographs (A.\ Uomoto et al., in preparation), 
careful calibration procedures
\citep{fukugita96,hogg01,smith02}, and an efficient series 
of data reduction and targeting pipelines 
\citep{lupton01,stoughton02,strauss02,blanton03a,pier03}.
Recently \citeauthor{blanton04a} (\citeyear{blanton04a}, hereafter 
B04) have searched the SDSS Second Data Release (DR2; \citealt{abazajian04}) 
to identify a population of extremely low luminosity field galaxies.  
With absolute magnitudes $M_r \sim -13$ to $-16$ 
(for the Hubble parameter $h\equiv H_0/100\hubunits=0.7$ adopted
throughout this paper), this 
population represents a range of luminosities that has previously been
accessible to systematic study 
only in the Local Group and in nearby groups and clusters.\footnote{The
field galaxy survey of \cite{schombert97}, based on visual inspection of
POSS-II plates and HI confirmation from Arecibo, includes some systems in
this luminosity range.}
We have obtained 
H$\alpha$ rotation curves of two of these objects, using the same
methods that we are using for a study of the Tully-Fisher (\citeyear{tul77}, 
hereafter TF) relation of more luminous SDSS galaxies 
(J. Pizagno et al., in preparation, hereafter P04).
Both objects appear to be ``field'' galaxies rather than satellites of
brighter systems.
Although the sample is small and the 
data quality limited, these measurements provide one of the 
first insights into the dynamical properties of this 
new population.

\section{Observations}
The galaxies in this paper were selected from the B04 sample according to 
their availability during an observing run for the P04 project.  As detailed 
in B04 and \cite{blanton04b}, this low luminosity galaxy sample has been
checked for contamination by double stars, errors in the photometric
pipeline's automated deblending, and other complications that are 
unimportant for the great majority of SDSS galaxies but have a 
significant impact for low luminosity and low redshift systems.
The two galaxies selected for follow-up observations 
were J123654.9+013654.2 and J091858.60+581407.7.
For brevity, we will hereafter refer to these as Galaxy 1 and Galaxy 2,
respectively.

Distances for galaxies in the B04 sample are estimated from their
redshifts assuming Hubble flow and a model for the local peculiar
field based on the IRAS 1.2-Jy redshift survey \citep{willick97}.
Details of this procedure are described by \cite{blanton04b}.
Because these galaxies have low redshifts (heliocentric values of 
$594\kms$ and $1157\kms$,
respectively), peculiar velocity uncertainties dominate the
distance uncertainties, which in turn dominate the uncertainties
in luminosity and dynamically inferred mass-to-light ratios in our
discussions below.  The \cite{blanton04b} procedure yields distances
of $9.9 \pm 2.9 \mpc$ and $20.0 \pm 2.2\mpc$ to Galaxies 1 and 2,
for $h=0.7$.
The $1\sigma$ error bars are derived from the peculiar velocity
probability function, which includes a $150\kms$ local dispersion
around the smooth velocity field predicted from the IRAS 1.2-Jy galaxies.
The corresponding Petrosian $r$-band absolute magnitudes are
$\Mr=-13.9^{+0.8}_{-0.5}$ and
$\Mr=-14.7^{+0.3}_{-0.2}$ (again for $h=0.7$).
Neither galaxy has a bright
neighbor ($\Mr < -19.25$) within 0.7 Mpc projected separation and 
$1000\kms$ redshift separation.
Galaxy 1 is close to NGC 4536 in projection, but the redshift difference,
$594\kms$ vs.\ $1808\kms$ \citep{grogin98}, makes a physical association 
unlikely.  We conclude that both of these objects are ``field'' galaxies
rather than satellites of brighter parents.

Table~1 lists the magnitudes, redshifts, and distances
for Galaxies 1 and 2 along with the
$g-r$ colors and axis ratios, exponential disk scale lengths, and
rotation speed parameters described below.
The top panels of Figure~1 show $i$-band images from the SDSS.
A search through the NASA Extragalactic Database (NED)
shows an HI observation (from \citealt{mat00})
for the galaxy FGC 1475, which can be identified as our Galaxy 1
on the basis of angular
proximity ($12''$ difference between the NED and SDSS positions), 
similar axis ratios, 
and similarity between the \cite{mat00} and SDSS redshifts.

Spectroscopic observations were made with the CCDS long-slit spectrograph, 
set up to observe redshifted H$\alpha$, at the 2.4 meter MDM 
telescope during the night of 13 April 2004.  
We used a $2''$ slit width with a 
600 lines/mm grating in the second order producing a 
0.41 \AA/pixel dispersion, and a
$0.41\,''/{\rm pixel}$ spatial scale.
The seeing during the night was $1.4-2''$.

The H$\alpha$ emission lines typically had a total 
signal-to-noise ratio of $6-15$, with an 
intensity weighted velocity centroid uncertainty of $\sim$ 
8 km/sec.  The dispersion axis was aligned to be perpendicular to the 
columns of the CCD to an accuracy of 0.1\AA, as judged from the 
telluric lines.  This alignment ensures that the \halpha\ recession velocity
is measured to the accuracy 
of the intensity weighted velocity centroids.
The details 
of the data reduction and rotation curve extraction can be found in P04.  

The bottom panels of Figure 1 show the linearized and 
flat-fielded spectra of both galaxies.  
Figure 2 presents the extracted rotation curve measurements as
observed velocity relative to the continuum center vs.\ position 
along the slit in arc-seconds.  
Following \cite{cou97} and P04, we fit the rotation curve measurements
with an arc-tangent function, which has a minimal number of 
free parameters while still describing the global 
shape of typical galaxy rotation curves quite well.
Specifically, we use a 
Levenburg-Marquardt $\chi^2$ minimization routine \citep{pre92} to 
fit the data with the functional form
\begin{equation}
V(r) = \Vcom + \frac{2}{\pi} \Vcirc {\rm arctan}\left(\frac{r-r_0}{r_t}\right),
\end{equation}
where $\Vcom$ is the central velocity,
$\Vcirc$ is the asymptotic 
circular velocity, $r$ is the position along the slit, $r_0$ is the 
center of the rotation curve (where $V=V_0$), 
and $r_t$ is the turnover radius at which the 
rotation curve begins to flatten.
The parameter uncertainties are derived from 
the covariance matrix returned by the Levenburg-Marquardt routine.
For Galaxy 1, we allow all four parameters ($\Vcom$, $\Vcirc$, $r_0$, $r_t$)
to vary.  For Galaxy 2, the best-fit arc-tangent parameters have very 
large $\Vcirc$ and $r_t$, so that they effectively describe a straight 
line over the region covered by the data points.  We therefore fit a 
straight line to the rotation curve instead of an arc-tangent function,
simplifying the 
determination of best-fit parameters and uncertainties.  Since Galaxy 
2's rotation curve is more extended on one side, we 
fix the line's intercept such that the line goes through the 
location of the conintuum center at the center of mass velocity.

Figure~2 shows the arc-tangent and linear fits to the data points as smooth solid curves.
For Galaxy 1, the outermost data points probe the turnover region of the
rotation curve, while for Galaxy 2 they are still linearly 
rising.  Clearly these data do not yield good constraints on the 
asymptotic circular velocity of a flat rotation curve.
However, we find in P04 that the velocity $\Vend$ defined by the value
of the arc-tangent fit at the radius of the outermost data point often
provides a useful measurement of rotation speed even when the observed rotation
curve is still rising at this radius.  In particular, we find that if
we use $\Vend$ as the measure of rotation speed in the TF relation, 
then the observed TF scatter does not increase significantly
when we include galaxies that have rising rotation curves
in the sample.  While the value of $\Vend$ is a lower limit to the
asymptotic circular velocity, it can be much more robustly measured
than $\Vcirc$ from truncated rotation curves like those in Figure~2.  

We correct $\Vend$ for inclination by using the GALFIT program
\citep{pen02} to fit inclined exponential disks to the $i$-band 
galaxy images shown in Figure~1.
The disk axis ratios and exponential scale lengths are listed in 
Table~1; formal uncertainties are $\sim 0.01$ in $b/a$ and $\sim 0.3''$
in $\Rexp$, though the systematic errors associated with assuming
an exponential disk model are probably larger.
The inclination corrected velocity is
\begin{equation}
\Vend^i = \Vend\left(\frac{1-b^2/a^2}{1-0.19^2}\right)^{-1/2}~,
\end{equation}
where 0.19 is the assumed intrinsic axis ratio for an edge-on disk and 
$b/a$ is the measured $i$-band axis ratio. Different ranges for the 
intrinsic axis ratio vary by 0.10 to 0.20 depending on galaxy type 
\citep{hay84}.  We chose 0.19, typical for spiral galaxies, and note 
that the range in intrinsic axis ratios causes a small variation 
(typically $\sim$1$\kms$) in $\Vend^i$.  
GALFIT yields a position angle for Galaxy 2 that differs from the SDSS
DR2 value by ten degrees, consistent with the visual impression of slight
misalignment in Figure~1.  We therefore apply a small ($\sim 1\kms$)
correction to the measured $\Vend$ 
assuming a tilted-ring model as described by 
\citeauthor{bea01} (\citeyear{bea01}; see their eq.~3).
Table~1 lists inclination-corrected values $\Vend^i$ including this
slit misalignment correction for Galaxy 2.

\cite{mat00} 
report an HI velocity width for Galaxy 1 of $W_{50,c} = 94\kms$,
where the subscript $c$ indicates a correction for instrumental resolution.
They do not give an error bar on $W_{50,c}$, though the total signal-to-noise
ratio of the HI line is 4.8, so the uncertainty in the width is probably
not negligible.
\cite{kan02} compare HI linewidths to optically measured, maximum
rotation speeds and report a correlation 
\begin{equation}
W_{50} = 19(\pm6) + 0.90(\pm0.03)(2\Vmax)~,
\end{equation}
in $\kms$ units.
Equation~(3) predicts $\Vmax=42 \pm 5 \kms$, compared to our
measured value of $\Vend=34.8\pm 3.8\kms$.
(Note that the inclination correction is negligible for this galaxy.)
The two measurements thus appear to be consistent within the observational
errors (allowing for a few percent uncertainty in the measured $W_{50,c}$),
though the somewhat higher value inferred from the HI data could reflect
a continuing rise of the rotation curve beyond the radius probed
by our \halpha\ measurements.

\section{Discussion}

How do our measured rotation speeds compare to expectations based on 
the TF relation defined by more luminous galaxies?  In P04, we use
a sample of 170 galaxies with $-22 < \Mr < -18$ to 
measure the forward ($M$ vs. $\log V$) and inverse ($\log V$ vs. $M$)
TF relations in the SDSS bands.
For the $r$-band inverse relation, we fit
\begin{equation}
\eta=a(\Mr-M_{r,0})+b
\end{equation}
with $\eta \equiv \log_{10}\Vend^i$
and the constant $M_{r,0}=-20.873$ chosen to yield
minimal correlation between the errors in $a$ and $b$.
We find $a=-0.143 \pm0.018$ and $b=2.202 \pm0.004$, yielding predicted values
of $\Vend^i$ for Galaxies 1 and 2 of $V_{\rm pred}=16.0^{+6.1}_{-5.4} \kms$ and
$V_{\rm pred}=20.9^{+6.2}_{-5.2} \kms$, respectively.  
The uncertainties in $\Mr$ and in the TF slope both contribute significantly
to the errors on $\Vpred$, and both contributions are non-linear.
We have computed upper $1\sigma$ error bars by separately varying
$\Mr$ by $-1\sigma$ and $a$ by $+1\sigma$ and adding the corresponding
values of $\delta \Vpred$ in quadrature; we follow an analogous procedure
for the lower $1\sigma$ error bar.  The predicted rotation
speeds are well below the measured values of $34.8 \pm 3.8 \kms$ and
$30.9 \pm 7.2\kms$, though marginally compatible in the latter case
because of the large observational errors.
Dotted curves in Figure~2 show rotation
curves with the same parameters as the solid curves but scaled by
$(V_{\rm pred}/\Vend^i)$.  Despite the uncertainties in the measurements,
it is clear that the galaxies are rotating substantially faster than
predicted by an extrapolation of the inverse TF relation to these
faint magnitudes.  To match the observed rotation speeds while maintaining
the P04 normalization at $\Mr \approx -21$ would require an inverse
TF slope of $a \approx -0.1$, compared to the slope of $-0.143$ measured
in the bright galaxy regime.

We can use our measurements to estimate dynamical mass-to-light ratios
within the radius of the outermost \halpha\ data point in each galaxy.
We use the simple mass estimator 
\begin{equation}
M=\frac{(\Vend^i)^2\Rend}{G}~,
\end{equation}
ignoring possible corrections for asymmetric drift or non-circular
motions, which would be small compared to our observational error bars.
The values of $\Rend$ are 19 and 11 arc-seconds for Galaxies 1 and 2,
respectively, corresponding to 1.7 and 1.8 times the exponential scale
lengths measured by GALFIT.  At our estimated distances of $9.9 \pm 2.9$ Mpc
and $20 \pm 2.2$ Mpc, the physical values of $\Rend$ are $0.91 \pm 0.27$ kpc
(Galaxy 1) and $1.07 \pm 0.12$ kpc (Galaxy 2).
We measure the fraction of $r$-band light within $\Rend$ for each galaxy
and correct the values of $\Mr$ in Table~1 accordingly, by 0.24 and 0.14
magnitudes.  We do not apply any internal extinction corrections.
Adopting $M_{r,\odot}=4.67$, we find
$M/L=12.6^{+4.7}_{-4.5}\, M_\odot/L_\odot$ for Galaxy 1 and 
$4.8_{-2.1}^{+2.5}\, M_\odot/L_\odot$ for Galaxy 2.
Because $M/L \propto \Vend^2/d$ and the fractional uncertainties in
these quantities are substantial, we compute the upper $1\sigma$ error
bar by separately varying $d$ by $-1\sigma$ and $\Vend$ by $+1\sigma$
and adding the two changes $\delta M$ in quadrature, and we follow
an analogous procedure for the lower $1\sigma$ error bar.
The $M/L$ value for Galaxy 1 is much higher than the
value $\sim 1.3 M_\odot/L_\odot$ expected for a stellar population
with the observed $g-r$ color of the galaxy \citep{bell03}, so this
system is strongly dominated by dark matter within $\Rend$.
The case for dark matter domination in Galaxy 2 is less clear because
of the large uncertainty in $M/L$, though the central value
again corresponds to a large ratio of dark matter to stellar mass
within $\Rend$.  (The \citeauthor{bell03} [\citeyear{bell03}] models
imply a stellar mass-to-light ratio of $\sim 0.9 M_\odot/L_\odot$ for
Galaxy 2's $g-r$ color.)
While all disk galaxies become dark matter
dominated at sufficiently large radii, these low luminosity systems
appear to have high dark matter fractions even within two disk
scale lengths.

To put our measurements in the context of previous results, 
Figure~3 plots inclination corrected circular velocities against
$B$-band absolute magnitude for our galaxies (filled circles with
error bars) and for galaxies with HI line widths and 
$M_B \ga -15$ from \citeauthor{cot00}
(\citeyear{cot00}; open squares), \citeauthor{beg04} 
(\citeyear{beg04}; open circles), and HI line widths 
from \citeauthor{carignan88}
(\citeyear{carignan88}; open triangle).  For our galaxies, we use
$\Vend^i$ as the indicator of circular velocity, and 
we convert the (AB) $g$-band
luminosity to the (Vega) $B$-band luminosity using the galaxy $g-r$ color and
$M_B = M_g + 0.365 + 0.46[(g-r) - 0.78]$, obtained using the 
$K$-correction code of \cite{blanton03b}.
For \cite{cot00} we use the value of $V_c$ 
listed in their tables, and for \cite{beg04} and \cite{carignan88}
we estimate $V_c$ visually
from the plotted rotation curves.  We have not attempted to put 
observational error bars on the literature data points because this
would require a detailed assessment of the uncertainties in the
distance measurements, which use a somewhat different method in each case.
\cite{cot00} observe galaxies in the Sculptor and Centaurus A groups
and assign group distances based on a variety of indicators
(see \citealt{cote97} and references therein).
\cite{beg04} assign both of their galaxies to the NGC 4696 group
on the basis of proximity and adopt a distance based on the
brightest stars \citep{huchtmeier00}.  \cite{carignan88} use a combination
of group assignment and brightest stars to infer the distance of DDO 154.

The solid line in Figure~3 shows the $B$-band inverse TF relation 
extrapolated to the low luminosity regime.  
We convert the $g$-band relation of P04 to $B$-band using $M_B=M_g+0.19$,
appropriate for a galaxy with $g-r=0.4$.
Dotted lines show relations with the slope varied
by $\pm 1\sigma$.  For bright galaxies, the intrinsic scatter about
the mean relation is $0.07\,$dex in $\log_{10}\Vend^i$ (P04),
smaller than the $0.1\,$dex observational error bar on Galaxy 1.
As noted earlier (on the basis of $r$-band data), Galaxy 1 is rotating
substantially faster than the extrapolated TF relation predicts, while
the large error bar on $\Vend$ leaves Galaxy 2 marginally 
consistent with the TF extrapolation.  The rotation speed of Galaxy 1
appears reasonably in line with that of the other dwarf galaxies,
while the rotation speed of Galaxy 2 is noticeably low, perhaps
because the optical rotation curve is still rising steadily at our
outermost data point.  All of these systems are rotating faster than
the TF extrapolation predicts; equivalently, they are fainter than
predicted given their rotation speeds.  

\cite{mcgaugh00} show that low luminosity galaxies are fainter
than predicted by the (forward)
TF relation defined by bright galaxies, consistent with the result
shown in Figure~3.  However, faint galaxies are often gas rich, and they
show that adding the gas masses inferred from HI observations to the stellar
masses leads to a linear relation between $\log M_{\rm bar}$ and 
$\log V_c$ over four decades in baryonic mass $M_{\rm bar}$, extending down
to a few $\times 10^7 M_\odot$.  For our Galaxy 1, \cite{mat00}
report an HI flux integral of 1.24 Jy$\kms$, which implies an HI mass to
optical luminosity ratio of $\sim 0.95 M_\odot/L_\odot$ in $r$-band, or
roughly equal mass in neutral gas and stars.  This ratio is below the median
of $\sim 2M_\odot/L_\odot$ found (in $V$-band) by \cite{pildis97}, but 
within the range spanned by their data.  Simply doubling the luminosity
of Galaxy 1 and thus shifting it rightwards by 0.75 mag in Figure~3 would
not move it onto the P04 inverse TF relation.  However, \cite{mcgaugh00}
find a steep slope for the (forward) baryonic mass TF relation, and the
baryonic mass of $7.5\times 10^7 M_\odot$ inferred for Galaxy 1 at a 
distance of 9.9 Mpc agrees respectably with the value 
$M_{\rm bar} = 30.5 (h/0.7)^{-2} (V_c/\kms)^4 M_\odot = 5.3 \times 10^7M_\odot$
predicted by their fitted relation for $V_c=35\kms$, given the substantial
error bars on both the predicted and measured values. 

The luminosity-velocity relations and mass-to-light ratios of low 
luminosity galaxies offer clues to the role of supernovae and 
photoionization in regulating star formation, since these feedback
effects are generally expected to be stronger in lower mass halos
(see, e.g., \citealt{dekel86,quinn96,thoul96,benson02,somerville02}).
The influence of feedback may be different in galaxies that are 
central objects of their parent dark matter halos and in galaxies that
are satellites in more massive halos --- semi-analytic models 
and hydrodynamic simulations predict systematically different properties 
for these two populations
\citep{kauffmann93,cole94,somerville99,berlind03,zheng04}.
Most of the very low luminosity galaxies studied to date have been
discovered or investigated because they are members of groups or
clusters defined by brighter galaxies.  
The two galaxies studied in this paper are not satellites of bright
neighbors, but their properties are roughly in line with those of
satellite dwarfs.
The B04 sample drawn from the
SDSS provides the opportunity to study such faint galaxies in detail over the 
full range of environments in which they appear.  Our present
investigation represents a start on this broader program.

\acknowledgments

JP and DW acknowledge support from NSF Grants AST-0098584 and
AST-0407125.  We thank Eric F. Bell for useful comments.  
Funding for the creation and distribution of the SDSS Archive has been 
provided by the Alfred P. Sloan Foundation, the Participating Institutions, 
the National Aeronautics and Space Administration, the National Science 
Foundation, the U.S. Department of Energy, the Japanese Monbukagakusho, 
and the Max Planck Society. The SDSS Web site is http://www.sdss.org/.

The SDSS is managed by the Astrophysical Research Consortium (ARC) for 
the Participating Institutions. The Participating Institutions are 
The University of Chicago, Fermilab, the Institute for Advanced Study, 
the Japan Participation Group, The Johns Hopkins University, 
Korean Scientist Group, Los Alamos National Laboratory, the 
Max-Planck-Institute for Astronomy (MPIA), the Max-Planck-Institute 
for Astrophysics (MPA), New Mexico State University, University of Pittsburgh, 
Princeton University, the United States Naval Observatory, and the 
University of Washington.

\clearpage

\begin{deluxetable}{cccccccccc}
\tabletypesize{\scriptsize}
\tablecaption{Galaxy Properties\label{tbl-1}}
\tablewidth{0pt}
\tablehead{
\colhead{Galaxy} & 
\colhead{$r$} & 
\colhead{$g-r$} & 
\colhead{$cz$} & 
\colhead{$d\,(\mpc)$} & 
\colhead{$\Mr$} & 
\colhead{$\Rexp\,('')$} & 
\colhead{b/a} &
\colhead{$\Vend$} &
\colhead{$\Vpred$} 
}
\startdata
J123654.9+013654.2 &
16.11 &  0.41 & 594 & $9.9 \pm 2.9$ & $-13.9^{+0.8}_{-0.5}$ & 11.2 & 0.20 & 
  $34.8\pm 3.8$ & $16.0^{+6.1}_{-5.4}$ \\
J091858.60+581407.7 &
16.84 &  0.31 & 1157 & $20.0 \pm 2.2$ & $-14.7^{+0.3}_{-0.2}$ & 6.2 & 0.40 & 
 $30.9\pm 7.2$ & $20.9^{+6.2}_{-5.2}$
\enddata
\tablecomments{Velocity units (for $cz$, $\Vend$, and $\Vpred$) are $\kms$.
Distances and absolute magnitudes are computed assuming
$H_0=70\hubunits$, using a model of the local peculiar velocity field
as described in the text.  All the magnitudes are corrected for Galactic 
foreground extinction.}
\end{deluxetable}

\clearpage

\begin{figure}
\epsscale{0.80}
\plotone{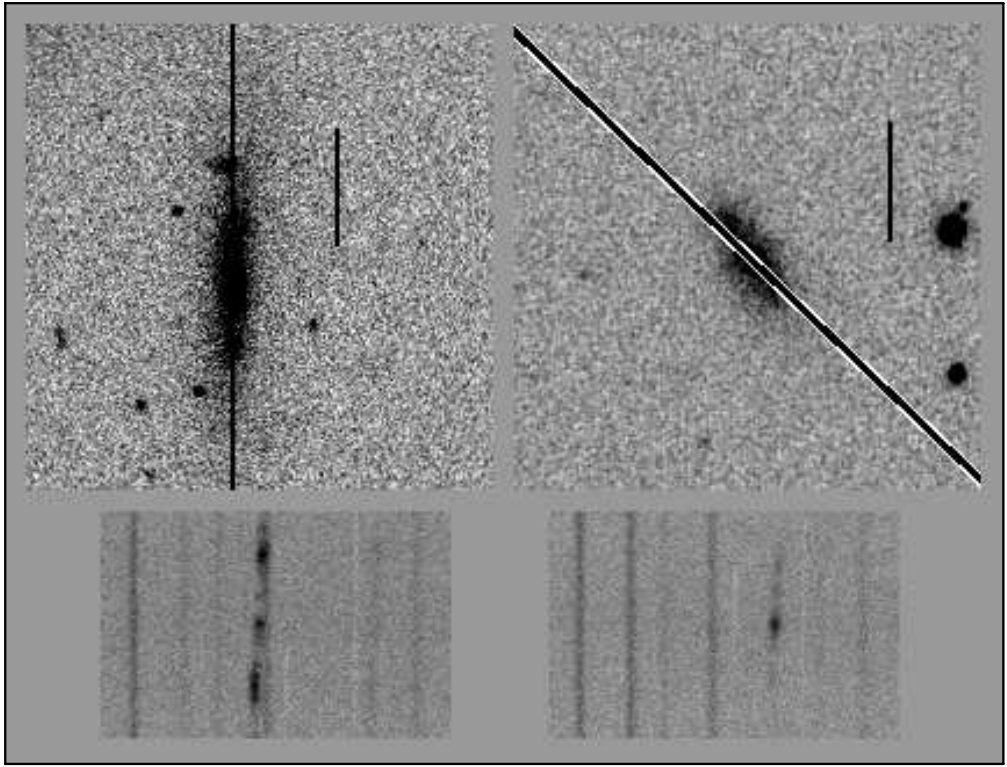}
\caption{Images and spectra of Galaxy 1 (left) and Galaxy 2 (right).
Upper panels show the SDSS $i$-band images with North down and East to the 
left.  The small vertical line has a length of 20 arc-seconds, and
the long line through each galaxy shows the slit position angle.
Lower panels show the flat-fielded, linearized spectra from MDM.
These panels are 40 arc-seconds high and 61 \AA\ wide, centered at
6582 \AA.
}
\end{figure}
\clearpage

\begin{figure}
\epsscale{.80}
\plotone{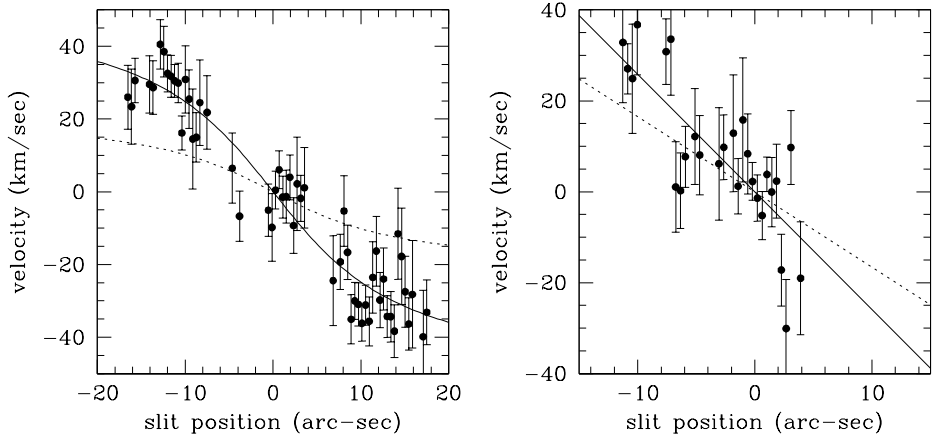}
\caption{Rotation curves of Galaxy 1 (left) and Galaxy 2 (right).
Points with error bars show the flux-weighted velocity centroid 
of the \halpha\ line as a function of position along the slit.
For Galaxy 1, the solid curve shows the arc-tangent 
fit to the data (eq.~1).  For Galaxy 2, we use a straight fit.  
Our measure of circular velocity is the value $\Vend$ of the
smooth fit at the radius of the outermost data point.
Dotted curves show the arc-tangent or linear fits scaled by
$\Vpred/\Vend$, where $\Vpred$ is the value predicted by
extrapolating the inverse TF relation of P04 to the absolute
magnitudes of these galaxies.  Note that these plots are not
corrected for inclination.
}
\end{figure}

\clearpage
\begin{figure}
\plotone{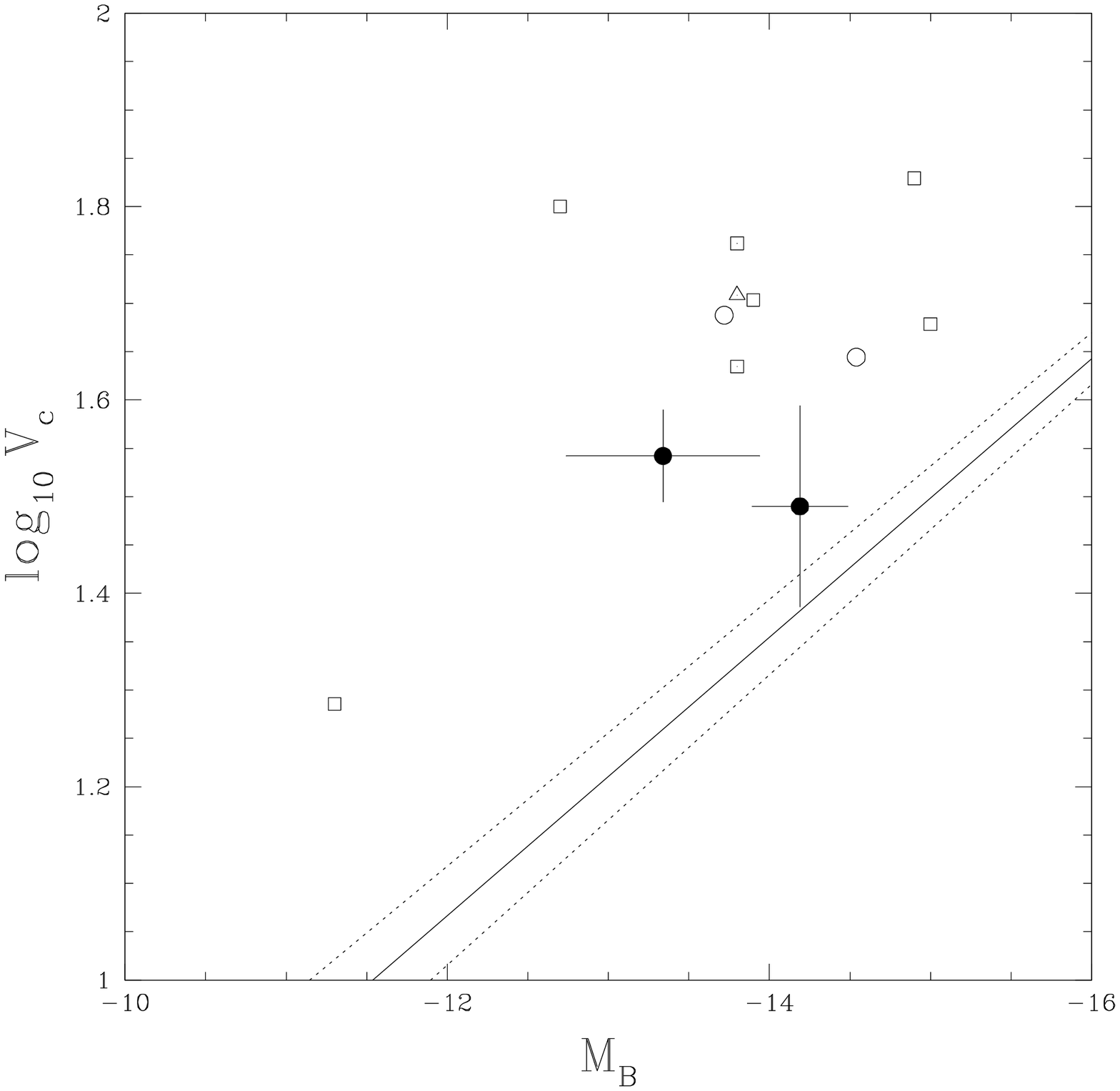}
\caption{
Inclination corrected circular velocity vs. $B$-band absolute magnitude
for our galaxies (filled circles with error bars) and for low luminosity
galaxies with HI data taken from 
\citeauthor{cot00} (\citeyear{cot00}, open squares),
\citeauthor{beg04} (\citeyear{beg04}, open circles), and
\citeauthor{carignan88} (\citeyear{carignan88}, open triangle).
The solid line shows the extrapolation of the $B$-band inverse TF relation 
to the faint galaxy regime.
The $B$-band inverse TF relation is found by converting the 
$g$-band inverse TF relation of P04.  
Dotted lines show this relation with the slope changed by $\pm 1\sigma$.
}
\end{figure} 

\end{document}